\documentclass[%
 amsmath,amssymb,
 aps,
prd,
floatfix,
onecolumn]{revtex4-2}

\usepackage{graphicx}
\usepackage{dcolumn}
\usepackage{bm}
\usepackage{tikz}
\usepackage{ulem}
\usepackage{subcaption}
\usepackage{hyperref}

\usetikzlibrary{decorations.pathmorphing,patterns}

\begin{document}


\title{Scattering off unstable states}%

\author{Francesco Giacosa}
 \email{francesco.giacosa@gmail.com}
\affiliation{Institute of Physics, Jan Kochanowski University, ul. Uniwersytecka 7, Kielce, Poland\\
Institute for Theoretical Physics, J. W. Goethe University, Max-von-Laue Str. 1, Frankfurt, Germany}
\author{Vanamali Shastry}%
 \email{vanamalishastry@gmail.com}
\affiliation{%
 Center for  Exploration  of  Energy  and  Matter, Indiana  University, Bloomington,  IN  47403,  USA\\
 Department of Physics, Indiana University, Bloomington, IN 47405, USA
 }%

\begin{abstract}
    Unstable states that live long enough may appear as in(out)going particles in scattering experiments. Yet, the standard QFT approach strictly applies only to fully stable asymptotic states. This is evident when scattering involving unstable particles develops a $t$-channel singularity at specific angles. We employ a finite-time formalism leading to analytic results without the singularity (even in the infinite-time limit), thus solving the problem at a phenomenological level. In turn, the approach also justifies treating long-lived particles, like weakly decaying pions, as stable during strong interactions.
\end{abstract}

\maketitle
\section{introduction}
The vast majority of particles listed in the Particle Data Group (PDG) compilation are unstable~\cite{ParticleDataGroup:2024cfk}. Among leptons, the muon and the tau particles decay weakly, and the neutrinos oscillate into each other. Among hadrons,  only the proton is currently known to be stable, while all other hadrons present in the PDG decay via weak, electromagnetic, or strong interactions.

If an unstable particle, elementary or not, lives long enough, it may be implemented as an incoming/outgoing state in scattering experiments. 
Typical examples are the $\mu^+ \mu^-$, pion-pion, pion-nucleon, and nucleon-nucleon processes.
With the advances in femtoscopy techniques, the scattering of additional unstable hadrons (including those that strongly decay) can be studied experimentally~\cite{Fabbietti:2020bfg}.

Nevertheless, the usually applied quantum field theoretical (QFT) formalism is rigorously defined only for asymptotically stable states. 
One may at first ignore this aspect and proceed with the textbook scattering of stable particles, e.g. Ref.~\cite{Peskin:1995ev}, arguing that it is applicable also for unstable states that live long enough. However, there are some cases where the standard approach fails badly because a divergence appears in the differential cross-section. This divergence is referred to as the `$t$-channel singularity' and was originally discussed in 1961 by Peierls~\cite{Peierls:1961zz} in the framework of the hadronic $N^*\pi$ scattering via an exchange of a (stable) nucleon $N$. This singularity also affects, e.g. $W^+ e^+$, $Z^0 e^-$, and $\mu^+ \mu^-$ scattering~\cite{Grzadkowski:2021kgi}.

In this work, we first recall how this singularity emerges in the conventional S-matrix approach, as well as why various previously proposed solutions cannot be regarded as conclusive (Sec. II). Next, in Sec. III we show how the QFT approach with finite time interval allows to study two-body scattering with initial/final unstable states and naturally avoids the singularity, even in the limit in which the time interval is sent to infinity (the cross-section simply vanishes in this case). Various discussions, including the comparison to three-body scattering and outlooks,
are contained in Sec. IV. Technical details are presented in the Appendix. 

\section{Conventional approach and t-channel singularity}
For simplicity, but with no loss of generality, let us consider a theory with 3 scalar fields, denoted with $A$, $B$, and $C$ with masses $m_{A,B,C}$. We choose $m_C \geq  m_B \geq m_A $.
In particular, we assume that $m_C >m_A+m_B$, rendering the decay $C \rightarrow AB$ kinematically possible. The corresponding interaction Lagrangian
\begin{equation}
    \mathcal{L} = gABC
\end{equation}
leads to the decay width
\begin{equation}
  \Gamma_{C\rightarrow AB}=\frac{g^2 |{\vec{k}_A}|}{8\pi m_C^2}  
\end{equation}
where $\vec{k}_A$ is the three-vector of the outgoing particle A.

Consider the scattering $AC \rightarrow AC$ via the exchange of $B$ as shown in Fig.~\ref{fig:tTree}.
The tree-level amplitude for this diagram is,
\begin{align}
i\mathcal{M} = i\mathcal{M}_t+i\mathcal{M}_s &= \frac{i (i g)^2}{t-m_B^2 + i\varepsilon}+\frac{i (i g)^2}{s-m_B^2 + i\varepsilon}
\text{ ,}
\label{ampl}
\end{align}
where $t= (p_C-k_A)^2$ and $s = (p_A +p_C)^2$ are the usual Mandelstam variables \footnote{Strictly speaking, the exchange channel is of the $u$-type, but the literature deals with $t$-channel singularity, see below. The scattering angle $\theta$ is chosen as the angle between $\vec{p}_C$ and $\vec{k}_A$, in agreement with the $t$-channel.}. For the in(out)going momenta, see Fig. 1. 
Using the standard QFT (sQFT) approach for asymptotic states and neglecting the fact that C is not stable, the differential and the total cross-section reads in the center-of-mass (c.o.m.) frame~\cite{Peskin:1995ev}:
\begin{equation}
    \frac{d\sigma}{d\Omega} = \frac{1}{16E_{in}^{2}}\frac{\left\vert i\mathcal{M}\right\vert ^{2}}%
{(2\pi)^{2}}
\text{  ,  } \sigma=
\int\mathrm{d\Omega}\frac{d\sigma}{d\Omega}=\int_{0}^{\pi}\mathrm{d\theta
}\frac{d\sigma}{d\Omega}2\pi\sin\theta \text{ ,}
\label{naive}
\end{equation}
where $E_{in}=\omega_{A}(\mathbf{p}_c) +\omega_{B}(\mathbf{p}_c) $ is the total initial energy with $\omega_{A,C}(\mathbf{p}) = \sqrt{\mathbf{p}^2+m_{A,C}^2}$ being the energy of $A(C)$.

\begin{figure}[t]
    \centering
    \begin{tikzpicture}[scale=1.]
        \draw (0,0) -- (1,1) -- (1,2) -- (0,3);
        \draw (1,1) -- (2,0);
        \draw (1,2) -- (2,3);
        \draw[double,thick] (0,3) -- (1,2);
        \draw[double,thick] (1,1) -- (2,0);
        \node at (-0.25,-0.1) {$A$};
        \node at (-0.25,3.2) {$C$};
        \node at (2.25,-0.1) {$C$};
        \node at (2.25,3.2) {$A$};
        \node at (1.4,1.5) {$B$};
        \draw[->] (0,0) -- (0.5,0.5);
        \draw[->,double] (0,3) -- (0.5,2.5);
        \draw[->] (1,2) -- (1.5,2.5);
        \draw[->,double] (1,1) -- (1.5,0.5);
        \draw[->] (1,2) -- (1,1.5);
        \filldraw[black] (1,1) circle (2pt);
        \filldraw[black] (1,2) circle (2pt);
        \node at (0.8,0.4) {$p_A$};
        \node at (0.4,2.2) {$p_C$};
        \node at (1.8,0.6) {$k_C$};
        \node at (1.4,2.8) {$k_A$};
        \node at (0.7,1.5) {$q_B$};

        \node at (1, -0.75) {(a)};

        \draw (3.5,0.5) -- (4.5,1.5) -- (6,1.5) -- (7,2.5);
        \draw[double, thick] (3.5,2.5) -- (4.5, 1.5);
        \draw[double, thick] (6,1.5) -- (7,0.5);
        \draw[double,->] (3.5,2.5) -- (4.,2.);
        \draw[double,->] (6,1.5) -- (6.5,1.);
        \draw[->] (3.5,.5) -- (4.,1.);
        \draw[->] (6,1.5) -- (6.5,2.);
        \filldraw[black] (4.5,1.5) circle (2pt);
        \filldraw[black] (6,1.5) circle (2pt);
        \node at (3.25,0.25) {$A$};
        \node at (3.25,2.75) {$C$};
        \node at (7.25,0.25) {$C$};
        \node at (7.25,2.75) {$A$};
        \node at (5.25,1.25) {$B$};
        
        \node at (4.3,0.75) {$p_A$};
        \node at (4.3,2.25) {$p_C$};
        \node at (6.25,0.75) {$k_C$};
        \node at (6.25,2.25) {$k_A$};
        \node at (5.3,1.75) {$q_B$};

        \node at (5.25,-0.75) {(b)};
    \end{tikzpicture}
    \caption{(a) The $t$-channel and (b) the $s$-channel diagrams representing the $CA\to AC$ scattering.}
    \label{fig:tTree}
\end{figure}

\begin{figure}[h]
    \centering
    \begin{subfigure}[b]{0.45\linewidth}
        \centering
        \includegraphics[width=0.75\linewidth]{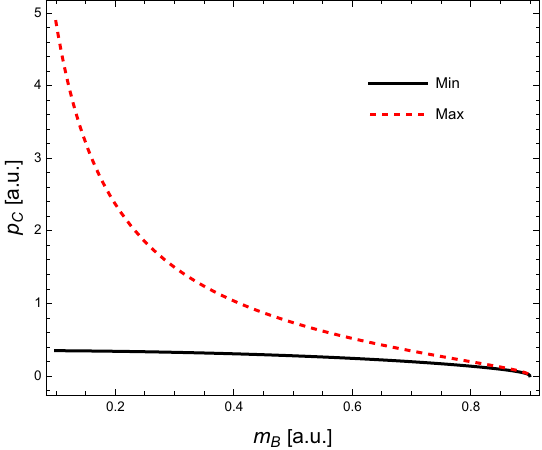}
        \caption{(a) $p_C$ vs. $m_B$.}
        \label{fig:a}
    \end{subfigure}
    \hspace{0.02\linewidth} 
    \begin{subfigure}[b]{0.45\linewidth}
        \centering
        \includegraphics[width=0.75\linewidth]{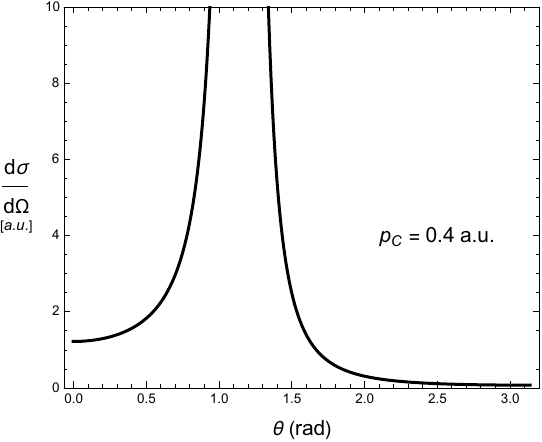}
        \caption{(c) $d\sigma/d\Omega$ vs. $\theta$.}
        \label{fig:c}
    \end{subfigure}

    \vspace{0.3cm}

    \begin{subfigure}[b]{0.45\linewidth}
        \centering
        \includegraphics[width=0.75\linewidth]{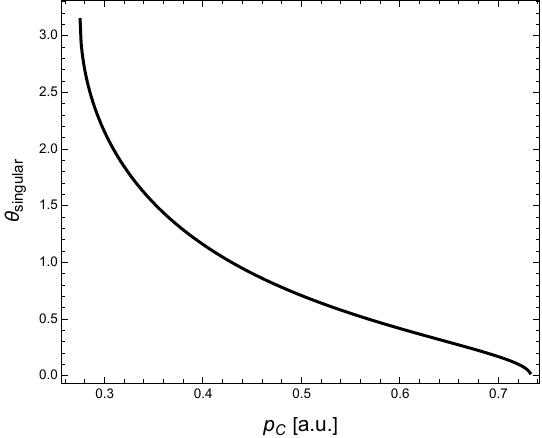}
        \caption{(b) $\theta_{\text{sing}}$ vs. $|\mathbf{p}_C|$.}
        \label{fig:b}
    \end{subfigure}
    \hspace{0.02\linewidth} 
    \begin{subfigure}[b]{0.45\linewidth}
        \centering
        \includegraphics[width=0.75\linewidth]{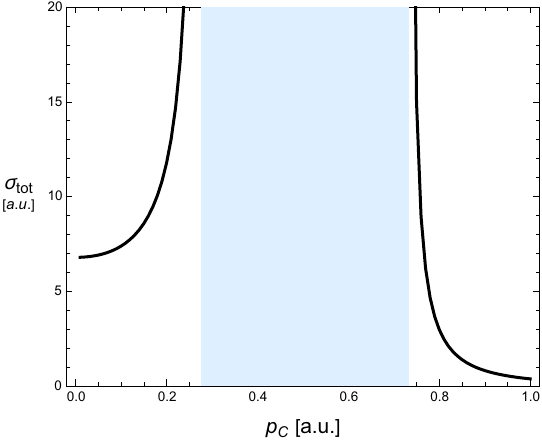}
        \caption{(d) $\sigma$ vs $|\mathbf{p}_C|$.}
        \label{fig:d}
    \end{subfigure}

    \caption{\raggedright Summary of plots for sQFT approach: 
    (a) Minimum and maximum values of $|\mathbf{p}_c|$ as a function of $m_B$ for $m_C=1$ and $m_A = 1/10$ [a.u.]. 
    (b) Singular angle $\theta_{\text{sing}}$ as a function of $|\mathbf{p}_C|$ for $m_B=0.5$ [a.u.].
    (c) Differential cross-section as a function of $\theta$ for $|\mathbf{p}_C| =0.4$ and $\Gamma_C =0.1$  [a.u.]. The singular angle is $\theta_{sing} = 1.15$ rad. 
    (d)  Total cross-section as a function of $p_C$: the band extends from $|\mathbf{p}_C|_{min} = 0.276$ [a.u.] to $|\mathbf{p}_C|_{max} = 0.733$ [a.u.], where the cross-section is not defined because of the singularity. }
    \label{fig:combined}
\end{figure}

Quite interestingly, if certain conditions are met, $t=m_B^2$ is possible, then a singularity in the $t$-channel amplitude of Eq.~\ref{ampl} appears.
More specifically, this is the case if three-momentum's length $\left\vert \mathbf{p}_{C}\right\vert$ takes the value between
\begin{equation}
\left\vert \mathbf{p}_{C}\right\vert _{\min}=\frac{\sqrt{m_{A}^{4}+m_{B}%
^{4}+m_{C}^{4}-2m_{A}^{2}m_{C}^{2}-2m_{B}^{2}m_{C}^{2}}}{2\sqrt{2m_{A}%
^{2}-m_{B}^{2}+2m_{C}^{2}}}%
\end{equation}
and
\begin{equation}
\left\vert \mathbf{p}_{C}\right\vert _{\max}=\frac{\sqrt{m_{A}^{4}+m_{B}%
^{4}+m_{C}^{4}-2m_{B}^{2}m_{C}^{2}-2m_{A}^{2}m_{C}^{2}}}{2m_{B}}%
\end{equation}
in the c.o.m frame. The value of the singular scattering angle $\theta_{sing}$ for which $t=m_B^2$ is given by
\begin{equation}
\cos\theta_{sing}=\frac{-(\omega_{C}(\mathbf{p}_{C})-\omega_{A}(\mathbf{p}_{C}%
))^{2}+m_{B}^{2}+2\left\vert \mathbf{p}_{C}\right\vert ^{2}}{2\left\vert
\mathbf{p}_{C}\right\vert ^{2}}
\text{ .}
\end{equation}
In particular, $\left\vert \mathbf{p}_{C}\right\vert _{\min}$ corresponds to $\theta_{sing} =\pi$ and $\left\vert \mathbf{p}_{C}\right\vert _{\max}$ to $\theta_{sing} =0$.
In Fig.~\ref{fig:combined}
we show the behavior of 
$\left\vert \mathbf{p}_{C}\right\vert _{\min} $, $\left\vert \mathbf{p}_{C}\right\vert _{\max}$ as functions of $m_B$, the singular angle $\theta$ as function of $\left\vert \mathbf{p}_{C}\right\vert$ for a given choice of $m_{A,B,C}$, and the `naive' differential and total cross-section in the standard QFT approach.  Whenever the singularity is present, the total cross-section diverges. 

\captionsetup[subfigure]{labelformat=empty}

Note, for the case $m_C < m_A+m_B$, the particle $C$ is stable, thus, no $t$-singularity appears. However, if $m_B > m_A+m_C$, a singularity appears at tree-level in the $s$ channel because $s=m_B^2$ is kinematically possible. This singularity is easily overcome by noticing that $B\rightarrow AC$ is now open; then, resumming the propagator of the $B$ particle amounts to the replacement $s - m_B^2 \rightarrow s - m_B^2 + im_B\Gamma_B $. Can a similar resummation solve the problem of the $t$-channel singularity occurring when $m_C > m_A+m_B$? This is not the case because the particle $B$ is truly stable. Hence, the pole of its propagator is physical and persists at any order. 

In order to understand better the singularity, it is useful to resort to an idealized case with a flux $j_C$ of $C$-particle and a single particle $A$ at rest. The number of scattered particles per second in the solid angle $d\Omega$ is $dN_{\text{scattered}} = j_C(d \sigma /d \Omega)d\Omega$. Thus, $dN_{\text{scattered}}$ diverges when $\theta$ tends to the singular value. This result is troubling and differs from Coulomb scattering, where the divergence takes place for $\theta \rightarrow 0$, meaning an arbitrarily imperceptible deviation of the scattered particles.

Various solutions to the $t$-channel singularity puzzle have been proposed.

1) A solution proposed in Ref.~\cite{Melnikov:1996iu} and further elaborated in Ref.~\cite{Dams:2002uy} takes into account the finite size of the beam, which cures the divergences. Following the heuristic discussion above, if the flux of $C$ has a limited section, the number of incident particles per unit time is finite and the cross-section cannot diverge. Interestingly, the reduction of the cross-section due to beam-size effect has been measured in Bremsstrahlung processes~\cite{Blinov:1982vp}. 

2) The exchanged particle $B$ has a finite width because it is not really stable: $m_B^2 \rightarrow m_B^2-im_B \Gamma_B$. In the case of $N^* \pi$ scattering, when the neutron is considered as the exchanged particle, the infinity is cured by the fact that the neutron has a nonzero width. Yet, since it is extremely small, the cross-section is very large. Moreover, this solution does not apply when the particle is stable, as in the case with an exchanged proton. There can be, however, other mechanisms that confer to the exchanged particle a finite width: these can be, for instance, in-medium effects~\cite{Grzadkowski:2021kgi,Iglicki:2022jjf}.

3) The particle $C$ is considered rightly as unstable. This point of view is based on the solution proposed by Ginzburg~\cite{Ginzburg:1995bc} and anticipated by Peierls~\cite{Peierls:1961zz}: $m_C^2 \rightarrow m_C^2-im_C \Gamma_C$. This simple change removes the divergence and generates `bumps' in the cross-section.

4) One may consider only stable states in the initial/final states. This amounts to considering the scattering process $AAB \to AAB$ containing $AC\to AC$ as a subprocess~\cite{Goebel:1964zz,Mai:2017vot,Jackura:2018xnx,Mikhasenko:2019vhk}. Alternative formalisms such as localized $S$-matrix theory have also been proposed to cure the divergence~\cite{Nowakowski:1993iu,Karamitros:2022nnh}.
\begin{figure}[h]
    \centering
    \begin{tikzpicture}
        \draw (0,0) -- (1,1) -- (1,2) -- (0,3);
        \draw (1,1) -- (2,0);
        \draw (1,2) -- (2,3);
        \draw[double,thick] (0,3) -- (1,2);
        \draw[double,thick] (1,1) -- (2,0);
        \draw (2,0) -- (3,0.5);
        \draw (2,0) -- (3,-0.5);
        \draw (0,3) -- (-1,3.5);
        \draw (0,3) -- (-1,2.5);
        \filldraw[black] (1,1) circle (2pt);
        \filldraw[black] (1,2) circle (2pt);
        \filldraw[black] (0,3) circle (2pt);
        \filldraw[black] (2,0) circle (2pt);
        \node at (-0.25,-0.1) {$A$};
        \node at (0.75,2.8) {$C$};
        \node at (1.25,0.25) {$C$};
        \node at (2.25,3.2) {$A$};
        \node at (1.4,1.5) {$B$};
        \node at (3.25,-0.5) {$A$};
        \node at (3.25,0.5) {$B$};
        \node at (-1.25,3.5) {$A$};
        \node at (-1.25,2.5) {$B$};
    \end{tikzpicture}
    \caption{One of the $3\to 3$ scattering diagrams that includes the $AC \to CA$ subprocess.}
    \label{fig:3to3}
\end{figure}

None of the above proposals fully resolves the issue. That is, in the first and second cases, the divergence appears again when the beam area increases or the width of the state $B$ (however obtained) decreases \footnote{In Ref.~\cite{Grzadkowski:2021kgi} the width of the state $B$ is generated by medium effects. The latter being {\it de facto} always present, the $t$-channel singularity is absent if the scattering occurs in the medium. In this respect, this type of solution is complete, but requires an external medium/field. Removing it, even in principle, restores the singularity.}. The third is rather an {\it ad hoc} description of the instability of $C$; although quite crude, it goes in the right direction, see below. 
The fourth point implies a rather complicated 3-body scattering, but the $t$-channel singularity seems to occur also in that case because the intermediate $B$ state can be generated on-shell \footnote{Namely, one may cut $AAB\rightarrow AAB$ diagram through an internal $B$ ({\it c.f.} Fig.~\ref{fig:3to3}), resulting in distinct sub-processes of the $AB\rightarrow AB$ types. A full treatment of this case is left for the future.}; moreover, one often deals with unstable states such as pions, kaons, muons, excited baryons, etc., as initial and final particles in two-body scattering (see the justification as a corollary of our treatment at the end of the paper). More details on the comparison with the three-body process of Fig. 3 are discussed in Sec. IV.

As emphasized by Collins~\cite{Collins:2019ozc}, the standard LSZ construction
relies on infinite-time limits and the corresponding operators do not create
strict one-particle states in a strong sense. Finite-time or smeared
formulations may therefore be more appropriate in situations involving
long-lived unstable particles. A specific finite-time construction is given below.

\section{QFT finite-time formalism}
In order to properly include the finite lifetime of $C$, we need to resort to QFT at finite times (see e.g. Refs.~\cite{Bernardini:1993qj,Blasone:2023brf,Blasone:2025hjw}).
Thus, at the leading order, we need to evaluate the scattering amplitude $i\mathcal{M}=\left\langle
f|S^{(2)}|in\right\rangle $ with the initial and final state given by
$\left\vert in\right\rangle =c_{\mathbf{p}_{C}}^{\dagger}a_{\mathbf{p}_{A}%
}^{\dagger}\left\vert 0\right\rangle $ and $\left\vert f\right\rangle
=c_{\mathbf{k}_{C}}^{\dagger}a_{\mathbf{k}_{A}}^{\dagger}\left\vert
0\right\rangle $ where $c_{\mathbf{p}}^{\dagger}$ and $a_{\mathbf{p}}%
^{\dagger}$ are the creation operators for $C$ and $A$, respectively.
The $S$-matrix at second order for a finite time interval $T$ reads
\begin{align}
S^{(2)}  &  =\frac{(ig)^{2}}{2!}\mathcal{T}\int_{-T/2}^{T/2}\mathrm{dt}%
_{1}\int\mathrm{d}^{3}\mathrm{x}_{1}\int_{-T/2}^{T/2}\mathrm{dt}_{2}%
\int\mathrm{d}^{3}\mathrm{x}_{2}
C(x_{1})B(x_{1})A(x_{1})C(x_{2})B(x_{2})A(x_{2})\text{ .}%
\end{align}
Next, we need to take into account the unstable nature of the
particle $C.\ $This implies that for the incoming $C$-particle created at
$t=-T/2$ and annihilated at $t=t_{1}\geq-T/2$, the usual phase is modified as:
\begin{equation}
e^{-i\omega_{C}(\mathbf{p}_{C})(t_{1}+T/2)}\rightarrow e^{^{-i\omega
_{C}(\mathbf{p}_{C})(t_{1}+T/2)}-\frac{\Gamma_{C}}{2\gamma_{C}(\mathbf{p}%
_{C})}(t_{1}+T/2)} \text{ .}
\label{Cunst}
\end{equation}
where, $\frac{\Gamma_{C}}{2\gamma_{C}(\mathbf{p}_{C})}$ represents the finite width correction to the energy, and $\gamma_C(\mathbf{p}_{C})$ is the corresponding Lorentz factor.
For $t_{1}=-T/2$ (initial time), there is no exponential suppression: there is simply no time for $C$ to decay. Yet, for growing $t_{1}$ the suppression is
taken into account. As discussed in detail in Refs.~\cite{Giacosa:2015mpm,Gavassino:2022ksl}, the $C$-lifetime $\gamma_{C}(\mathbf{p}%
_{C})/\Gamma_{c}$, although not exact, is typically an excellent approximation
for the decay law of a moving unstable particle. Namely, there are interesting
but very small deviations from the time-dilation expression, which, however, do not affect our discussion.\par

Similarly, for the outgoing $C$ state we replace: 
\begin{equation}
e^{i\omega_{C}(\mathbf{k}_{C})(t_{2}-T/2)}\rightarrow e^{i\omega
_{C}(\mathbf{k}_{C})(t_{2}-T/2)+\frac{\Gamma_{c}}{2\gamma_{C}(\mathbf{k}_{C}%
)}(t_{2}-T/2)}
\text{ ,}
\end{equation}
which describes the creation of the $C$ state at $t=t_{2}$ and its subsequent
measurement at $t=T/2\geq t_{2}.$ The term above contains the exponential
suppression due to the eventual decay of the $C$ state between creation and final time $T/2$.
The initial and final $C$-states can be regarded as Gamow states (e.g. Refs.
\cite{Civitarese:2004xbt,delaMadrid:2002cz} and refs. therein).\par

If applied to AB scattering, the formalism involving an unstable $C$ from the very onset can be applied to other processes as
well. For instance, for the $s$-channel $AB\rightarrow AB$ scattering, one obtains already at `tree-level' the resummed propagator
$1/(s-M^{2}+iM\Gamma_{C}),$ see e.g. \cite{Gegelia:2010nmt}. For more
technical details, see the Appendix.\par

The simple modification above has far-reaching consequences for $CA$ scattering. It cures the singularity for any value of $T$, including $T\to\infty$. The cross-section vanishes smoothly in this limit because the $C$ particle would decay during the process, implying that the limit $T \rightarrow \infty$ does not lead to the standard QFT result of Eq.~\ref{naive}. Schematically: 
\begin{equation}
    \text{for $\Gamma_C >0$: `$\sigma$ for large $T$}' \neq \text{`$\sigma$ in standard QFT'.}
    \nonumber
\end{equation}
Note, this outcome is \textit{not} analogous to the beam-size case. Both cure the divergences, but for the latter, the divergence reappears when the beam size is decreased .

The finite-$T$ cross-section reads:
\begin{equation}
\frac{d\sigma}{d\Omega}=\frac{1}{4\left\vert \mathbf{p}_{C}\right\vert E_{in}%
}\int_{0}^{\infty}\frac{\mathrm{d}\left\vert \mathbf{k}_{A}\right\vert
}{2\omega_{A}(\mathbf{k}_{A})2\omega_{C}(\mathbf{k}_{A})}\frac{\left\vert
\mathbf{k}_{A}\right\vert ^{2}}{(2\pi)^{3}}F[T,\left\vert \mathbf{k}%
_{A}\right\vert ,\theta] 
\text{ ,} 
\label{eq:diffcs}%
\end{equation}
where $F[T,\left\vert \mathbf{k}_{A}\right\vert ,\theta]$ reads:%

\captionsetup[subfigure]{labelformat=empty}

\begin{figure*}[htbp!]
    \centering
    \begin{subfigure}[b]{0.3\textwidth}
        \centering
        \includegraphics[width=\linewidth]{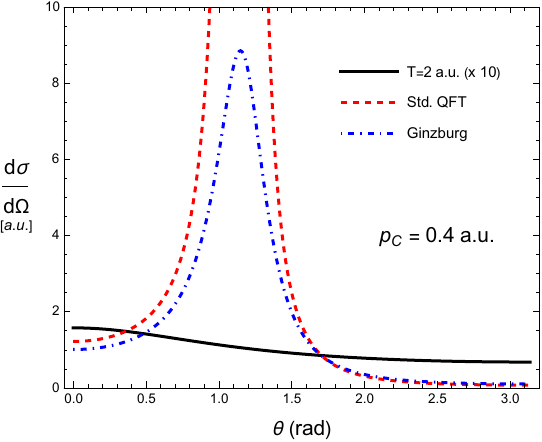}
        \caption{(a) $\frac{d\sigma}{d\Omega}$ vs.\ angle at $T = \tau/5$, standard QFT, and the Ginzburg solution~\cite{Ginzburg:1995bc}. Present work ×10 for visibility.}
        \label{fig:pAng}
    \end{subfigure}
    \hspace{0.02\linewidth}
    \begin{subfigure}[b]{0.3\textwidth}
        \centering
        \includegraphics[width=\linewidth]{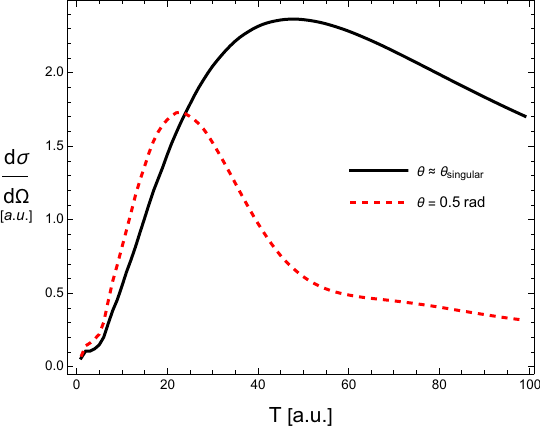}
        \caption{(b) $\frac{d\sigma}{d\Omega}$ vs. time at two angles — near  the naive standard QFT singularity (dot-dashed) and far from it (solid).}
        \label{fig:pTime}
    \end{subfigure}
    \hspace{0.02\linewidth}
    \begin{subfigure}[b]{0.3\textwidth}
        \centering
        \includegraphics[width=\linewidth]{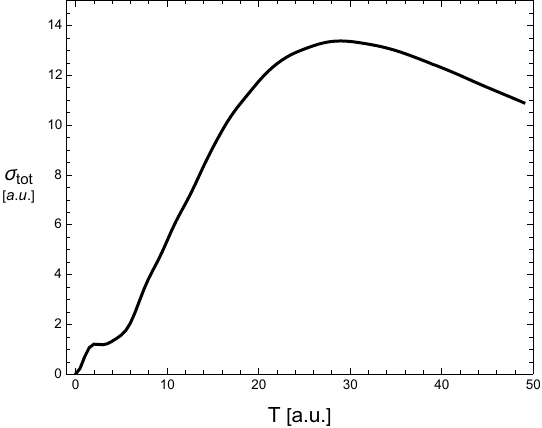}
        \caption{(c) Total cross section vs.\ $T$ at $p_C = 0.4$ a.u.}
        \label{fig:Totxsec}
    \end{subfigure}
    \caption{Overview of cross-section behavior.}
    \label{fig:cross_section_summary}
\end{figure*}

\begin{equation}
F[T,\left\vert \mathbf{k}_{A}\right\vert ,\theta]=\frac{1}{T}\left\vert
I_{F}^{(up)}+I_{F}^{(down)} + \text{$s$-channel}\right\vert ^{2}\text{ ,}\label{eq:Feq}%
\end{equation}
with (see the Appendix for details):
\begin{equation}
I_{F}^{(up)}=\frac{(ig)^{2}}{2E_{B}}\frac{e^{-i(P_{1}-P_{2})\frac{T}{2}%
}-e^{-i(P_{1}+P_{2}+2E_{B})\frac{T}{2}}+e^{i(P_{1}-P_{2})\frac{T}{2}}}{\left[
E_{B}+P_{1}\right]  \left[  E_{B}+P_{2}\right]  }\text{ ;}\label{eq:Ifu}
\end{equation}%
\begin{align}
I_{F}^{(down)} &  =\left[  \frac{(ig)^{2}}{2E_{B}}\frac{-e^{i(P_{1}%
+P_{2}-2E_{B})\frac{T}{2}}}{\left[  E_{B}-P_{1}\right]  \left[  E_{B}%
-P_{2}\right]  }\right.  
 +\frac{(ig)^{2}}{\left[  P_{1}-E_{B}\right]  \left[  P_{1}+E_{B}\right]
}\frac{e^{i(P_{2}-P_{1})\frac{T}{2}}}{P_{2}-P_{1}} \left.  +\frac{(ig)^{2}}{\left[  P_{2}-E_{B}\right]  \left[  P_{2}%
+E_{B}\right]  }\frac{e^{i(P_{1}-P_{2})\frac{T}{2}}}{P_{1}-P_{2}}\right]
\text{ .}\label{eq:Ifd}
\end{align}
Moreover, the poles are:
\begin{align}
P_{1}  &  =\omega_{C}(\mathbf{p}_{C})-\omega_{A}(\mathbf{k}_{A})-i\frac
{\Gamma_{C}}{2\gamma_{C}(\mathbf{p}_{C})}\text{ ,}\\
P_{2}  &  =-\omega_{A}(\mathbf{p}_{C})+\omega_{C}(\mathbf{k}_{A}%
)-i\frac{\Gamma_{C}}{2\gamma_{C}(\mathbf{k}_{A})}\text{ .}%
\end{align}
In the limiting case of $\Gamma_C = 0$,
\begin{equation}
F[T\rightarrow\infty,\left\vert \mathbf{k}_{A}\right\vert ,\theta]=\left(
2\pi\right)  \left\vert i\mathcal{M}\right\vert ^{2}\delta\left(
E_{in}-E_f\right) \text{ ,}
\end{equation}
with $E_f = \omega_{A}(\mathbf{k}_{A})+\omega_{C}(\mathbf{k}_{A})$. Then, the standard QFT result of Eq. (\ref{naive}) is recovered in this case, serving as a check of the formalism. Schematically: 
\begin{equation}
     \text{for $\Gamma_C  = 0$: `$\sigma$ for large $T$'} =  \text{`$\sigma$ in standard QFT.'}
    \nonumber
\end{equation}\par

 These analytic expressions constitute one of the main outcomes of this work because
 they allow for the calculation of $d\sigma/d\Omega$, and then of
$\sigma,$ in any desired case and for any value of the time interval $T$. The analogous $s$-channel expressions analogous to Eqs.~\ref{eq:Ifu},~\ref{eq:Ifd} are obtained by using $P_{s,1,2}$.

The angle dependence of the differential cross section obtained from Eq.~\ref{eq:diffcs}, combined with Eqs.~\ref{eq:Feq},~\ref{eq:Ifu},\ref{eq:Ifd}, are shown in Fig.~\ref{fig:pAng} for a specific choice $T/\tau= 1/5$. The singularity that is present in the standard QFT case (dashed curve) is cured when the finite lifetime of the unstable state is taken into account. 
As anticipated, the differential cross section vanishes when $T \gg \Gamma_C^{-1}$ since the state $C$ decays before interacting with $A$, see Eqs.~\ref{eq:Ifu}-\ref{eq:Ifd}. 
Thus, no divergence appears, independently of the choice of the time $T$, solving the $t$-channel problem. A peak (a la Ginzburg) in the cross-section is not visible for small $T$, but may emerge for intermediate/large ones. Yet, the numerical values are different: in no case does one observe a curve numerically similar to the Ginzburg proposal. 

The time-interval dependence of the differential cross section at two given illustrative angles is shown in Fig.~\ref{fig:pTime}. One sees a maximum after which it decreases smoothly to zero, as expected. Finally, the behavior of the total cross-section is shown in Fig.~\ref{fig:Totxsec}.

\section{Discussions and Conclusions}
In this concluding Section, we discuss some important points regarding the presented solution of the t-channel problem: the choice of the time interval $T$ in specific cases and the comparison to three-body scattering.

The value of the time interval $T$ is provided by the specific reaction. For scattering involving muons and/or other unstable particles, $T$ is the time between creation and detection of the states.  For femtoscopy, $T$ is rather short, of the order of fms. 
Within this context,  the experimental and theoretical study of $\phi K$ scattering is promising. Since the $\phi$ can decay into $\bar{K}K$, a situation very similar to our scalar model takes place. This is therefore a clean outlook of our approach, representing a direct application of the proposed method. 
Another similar femtoscopy study involves $K_S \pi$ scattering because of 
$K_S \rightarrow \pi \pi$, but in this case the decay is weak, thus strong and weak interactions mix, see below.

The presence of different interaction types that lead to different time scales has important consequences. As an illustrative example, we consider the interaction Lagrangian 
\begin{equation}
    \mathcal{L} = gABC + \lambda A^2 C^2 \text{ ,}
\end{equation}
 for which the amplitude for $AC \rightarrow AC$ is modified as 
 \begin{equation}
i\mathcal{M}_{new} =i\mathcal{M} + i4\lambda  \text{ .}
 \end{equation}
Suppose the $\lambda$-term is sufficiently large. In that case, one may indeed choose $T \ll \tau$ for which the cross-section contribution of the $g$-term is small, allowing to approximate the amplitude as $i\mathcal{M} \approx i4\lambda$ (without the $g$-term present), and use the standard QFT results with asymptotic states. Namely, Eqs.~\ref{eq:Ifu}-\ref{eq:Ifd} work {\it de facto} as Dirac-$\delta$ function for what concerns the $\lambda$-interaction, if the latter is large enough. In other words, $T$ can be chosen as being small for the $g$-term, but very large for $\lambda$-terms. The result is anything but naive: namely, if the $t$-singularity is present, there is `in principle' always a kinematic domain in which the allegedly dominating $\lambda$-term is subleading. That would invalidate the possibility of neglecting the fact that $C$ is not stable. The fact that one can indeed treat $C$ as stable if the $\lambda$-term dominates is provided by the finite-time QFT result discussed above. 

This discussion provides a formal explanation for a seemingly intuitive fact: in the case of scattering of states that live long enough (for instance, decaying weakly) one may treat a stronger interaction (such as the e.m. or the strong one, with much shorter time scales) as the only one present and apply the large-$T$ limit, thus in such cases the standard QFT approach can be used treating unstable states as stable ones: this is the case e.g.~pion-pion, pion-nucleon, and nucleon-nucleon scattering. As notable examples, the previously mentioned $K_S \pi$ and proton-neutron scattering would suffer from a $t$-channel singularity (when a pion or an electro-neutrino pair is exchanged, respectively). When strong interactions are considered, these processes must be, however, negligible, what our finite-$T$ formalism shows.

Finally, we come back to the three-body scattering $AAB\rightarrow AAB$ that
contains $CA\rightarrow CA$ as a subprocess (see Fig.~\ref{fig:3to3}). If a finite time
$T$ is kept, the internal $B$ cannot go on-shell, hence no singularity is present
in the three-body reaction rate. Yet, at odds with the two-body $CA$
scattering, this singularity appears back when $T\rightarrow\infty$ is taken,
since the intermediate particle may go on-shell for appropriate kinematical conditions.

Moreover, the emergence of the t-channel singularity is a general property of
three-body problems whenever the three-body interaction can be seen as two
subsequent two-body scattering processes (in our example, $AB\rightarrow AB$
two times).
Namely, for a process such as $AAB\to AAB$, the three-body amplitude contains
contributions that can be schematically written as
\begin{equation}
\mathcal{M}_{3}
\sim
\mathcal{M}_{2}(AB\to AB)\,
G_{\mathrm{B}}\,
\mathcal{M}_{2}(AB\to AB)\,,
\label{eq:seq3body}
\end{equation}
where $G_{B}$ denotes the propagation of intermediate $B$
particle.
Since $\mathcal{M}_{2}(AB\to AB)$ is a well-defined scattering amplitude between
stable asymptotic states, Eq.~\eqref{eq:seq3body} admits a physical
interpretation as two successive, time-separated two-body collisions.
In the limit of infinite interaction time, this leads to kinematical
singularities ($B$ can go on-shell) reflecting the fact that the second scattering can occur
arbitrarily long after the first.

Modern relativistic three-body formalisms therefore introduce a
divergence-free three-body amplitude by explicitly subtracting all such
sequential pairwise-scattering contributions, leaving a genuine three-body
interaction kernel.
In Ref.~\cite{Potapov:1977ux,Potapov:1977sr} a clear account of
the physics underlying this situation is investigated in a non-relativistic
approach, where it is dubbed as `apparent difficulty' (for another presentation
in terms of Coulomb interaction, see \cite{Rubin:1966zz}). It is shown that the physical
expression for the scattering rate\footnote{A universal definition of an
analogous of the cross-section for three-body scattering is not feasible,
since the specific invariant flux does not exist. Physically, for three
particles to interact, all three worldlines must meet in a very small
space--time region. The frequency of such coincidences depends on particle
densities, correlations, and beam geometry} reduces to a finite result that
contains the product of two standard scattering. In more recent relativistic
QFT approaches, such as Refs.~\cite{Hansen:2014eka,Baeza-Ballesteros:2023ljl},
the same issue is treated by introducing the divergence-free three-body
amplitude $\mathcal{M}_{\mathrm{df},3}$, obtained by explicitly subtracting
all contributions from repeated pairwise sequential two-to-two scatterings from the full three-body amplitude.

One might be tempted to apply an analogous reasoning to the two-body process
$CA\to CA$ for which the decay channel
$C\to AB$ is open.
Formally, the amplitude could be written in the form
\begin{equation}
\mathcal{M}(CA\to CA)
\;\sim\;
\mathcal{M}(C\to AB)\,
G_{B}\,
\mathcal{M}(AB\to C)\,,
\label{eq:fakefactor}
\end{equation}
suggesting an interpretation in terms of a decay followed by an `inverse decay'.
However, this factorization is fundamentally different from
Eq.~\eqref{eq:seq3body} and does not correspond to a sequence of distinct
scattering processes involving asymptotic states.
Namely, since $C$ is unstable, neither $C\to AB$ nor $AB\to C$ is a
well-defined $S$-matrix element between asymptotic states, as emphasized in
standard treatments of unstable particles \cite{Peskin:1995ev}.
Equation~\eqref{eq:fakefactor}, therefore, does not describe two time-separated
physical subprocesses.

Consequently, the singular behavior encountered in a naive treatment of
$CA\to CA$ with $C$ taken as an asymptotic particle does not originate from
double counting of sequential scattering processes and cannot be cured by a
subtraction analogous to the divergence-free three-body construction.
Rather, it signals the breakdown of the asymptotic-state assumption for the
unstable particle $C$ in the standard QFT (infinite time) approach..

Summarizing: (i) in the three-body case asymptotic states are implemented and
the time $T$ can be sent to infinity, but then the sequential scattering
processes should be properly treated/subtracted. (ii) In two-body scattering involving unstable particles, the initial and/or final
states do not need be true asymptotic states, as it is demanded in standard scattering theory. Nevertheless, the physical cross-section
can be consistently defined in the finite-$T$ formalism, where the unstable
particle is created and detected within finite time intervals.

In conclusion, in this work, we have shown that the scattering process is possible for initial/final unstable particles. The key point is to properly include the instability into the finite-time QFT formalism. 
In the proposed approach, no divergences appear, not even in the infinite-time limit. 

\bigskip 

\textbf{Acknowledgments} The authors thank S.~Mr\'owczy\'nski, A.~Pilloni, and A.~Szczepaniak for useful discussions.

\onecolumngrid
\appendix
\section{Details of the finite-time formalism}
\label{sec:app}
In this appendix, we introduce Gamow states for the unstable particles and present the derivation of the finite-time cross section. 
\subsection{Treatment of an unstable field}
The propagator of an unstable particle\ $C$ in the relativistic Breit-Winger
approximation reads
\begin{align}
\Delta_{C}(E,\mathbf{k}) &  \equiv \Delta_{C}(k)=\frac{i}{k^{2}-m_{C}^{2}%
+im_{C}\Gamma_{C}+i\varepsilon}\nonumber\\
&  \simeq\frac{i}{2\omega_{C}(\mathbf{k})}\left[  \frac{1}{E-\omega
_{C}(\mathbf{k})+i\frac{\Gamma_{C}}{2\gamma_{C}(\mathbf{k})}}-\frac{1}{E+\omega
_{C}(\mathbf{k})-i\frac{\Gamma_{C}}{2\gamma_{C}(\mathbf{k})}}\right]
\text{ .}
\end{align}
This
propagator is obtained by the field $C(x)$ that includes its finite decay width: 
\begin{equation}
C(x)=\int\frac{d^{3}k}{(2\pi)^{3}%
}\frac{1}{\sqrt{2\omega_{C}(\mathbf{k})}}\left(  c_{\mathbf{k}}e^{-ik\cdot
x-\frac{\Gamma_{C}(\mathbf{k})}{2}t}+c_{\mathbf{k}}^{\dagger}e^{ik\cdot
x-\frac{\Gamma_{C}(\mathbf{k})}{2}t}\right) \text{ ,}
\end{equation}
where $\Gamma_C(\mathbf{k})=\Gamma_C/\gamma(\mathbf{k})$. 
In turn, when $C(x)$ acts on the ket $c_{\mathbf{p}}^{\dagger}\left\vert
0\right\rangle $ one obtains
\begin{equation}
C(x)c_{\mathbf{p}}^{\dagger}\left\vert 0\right\rangle =\frac{e^{-ip\cdot
x- \frac{\Gamma_{C}(\mathbf{p})}{2}t}}
{\sqrt{2\omega_{C}(\mathbf{p})}%
}\left\vert 0\right\rangle + ... \text{ ,}%
\end{equation}
that is a decaying state created at time $t=0$ and probed at a later time
$t>0$. 
On the other hand, for the final state:
\begin{equation}
\left\langle 0\right\vert c_{\mathbf{p}}C(x)=\left\langle 0\right\vert
\frac{e^{ip\cdot x-\frac{\Gamma_{C}(\mathbf{p})}{2}t}}{\sqrt
{2\omega_{C}(\mathbf{p})}}
+ ... 
\text{ .}
\end{equation}
This can be interpreted as a state $C$ created at $-t<0$ and probed at $t=0.$
If we start at a time $t_{0}\neq0,$ we may consider the initial state
\begin{equation}
e^{i\omega_C(\mathbf{p})t_0+\frac{\Gamma_C(\mathbf{p})}{2}t_0}c_{\mathbf{p}%
}^{\dagger}\left\vert 0\right\rangle =\left\vert \mathbf{p};t_{0}%
\right\rangle
\end{equation}
that leads to
\begin{equation}
C(x)\left\vert \mathbf{p};t_{0}\right\rangle =\frac{e^{-i\omega
_{C}(\mathbf{p})(t-t_{0})+i\mathbf{p\cdot x}-\frac{\Gamma_{C}(\mathbf{p})}
{2}(t-t_{0})}}{\sqrt{2\omega_{C}(\mathbf{p})}}\left\vert 0\right\rangle 
+... \text{ .}
\end{equation}
In simple terms, this is a state created at $t_{0}$ and probed at a later time
$t>t_{0}.$ The ket $\left\vert \mathbf{p};t_{0}\right\rangle $ can be regarded
as a Gamow state. Thus, using the expressions above we can deal with scattering involving unstable states.
Finally, we report the propagator in position space:%
\begin{equation}
\left\langle 0\left\vert T\left\{  C(x)C(y)\right\}  \right\vert
0\right\rangle =\int\frac{d^{3}k}{(2\pi)^{3}}e^{i\mathbf{k\cdot}%
(\mathbf{x-y})}\Delta_{C}(t,\mathbf{k})=\int\frac{d^{4}k}{(2\pi)^{4}%
}e^{-i k\cdot(x-y)}\Delta_{C}(k)
\end{equation}
with
\begin{align}
\Delta_{C}(t,\mathbf{k}) &  =\int_{-\infty}^{+\infty}\frac{dE}{2\pi}%
e^{-iEt}\Delta_{C}(E,\mathbf{k})\nonumber\\
&  =\frac{1}{2\omega_{C}(\mathbf{k})}\left(  \theta(t)e^{-i\omega
_{C}(\mathbf{k})t-\frac{\Gamma_{C}(\mathbf{k})}{2}t} + \theta(-t)e^{i\omega
_{C}(\mathbf{k})t-\frac{\Gamma_{C}(\mathbf{k})}{2}\left\vert t\right\vert
}\right)
\text{.}
\end{align}
Future improvements should go beyond the relativistic Breit-Wigner approach. 

\subsection{Derivation of the cross section for finite $T$}
Here we explicitly derive the differential cross section formula given in Eq.~\ref{eq:diffcs}. The quantity $F[T,\left\vert \mathbf{k}_{A}\right\vert ,\theta]$ takes the form:
\begin{align}
F[T,\left\vert \mathbf{k}_{A}\right\vert ,\theta] &=
\frac{1}{T}\Big\vert
2\pi\int_{-\infty}^{+\infty}\mathrm{dq}^{0}\left(  i\mathcal{M}_t\right)
I_{1}(T,q^{0})I_{2}(T,q^{0}) + s\text{-channel}\Big\vert ^{2}
\label{eq:F1}
\text{ .}
\end{align}
The $\theta$ dependence is contained in the Feynman amplitude
$i\mathcal{M}$ of Eq. (\ref{ampl}). The functions $I_{1,2}(T,q^{0})$ read:
\begin{align}
I_{1}(T,q^{0})  &  =\frac{-ie^{-\frac{\Gamma_{C}}{2\gamma_{C}(\mathbf{p}_{C}%
)}\frac{T}{2}}}{2\pi}\frac{e^{-i(P_{1}-q^{0})T/2}-e^{i(P_{1}-q^{0})T/2}}%
{q^{0}-P_{1}}\text{ ,}\\
I_{2}(T,q^{0})  &  =\frac{ie^{-\frac{\Gamma_{C}}{2\gamma_{C}(\mathbf{k}_{A}%
)}\frac{T}{2}}}{2\pi}\frac{e^{-i(-P_{2}+q^{0})T/2}-e^{i(-P_{2}+q^{0})T/2}%
}{q^{0}-P_{2}}\text{ ,}%
\end{align}
where the following poles have been introduced:%
\begin{align}
P_{1}  &  =\omega_{C}(\mathbf{p}_{C})-\omega_{A}(\mathbf{k}_{A})-i\frac
{\Gamma_{C}}{2\gamma_{C}(\mathbf{p}_{C})}\text{ ,}\\
P_{2}  &  =-\omega_{A}(\mathbf{p}_{C})+\omega_{C}(\mathbf{k}_{A}%
)-i\frac{\Gamma_{C}}{2\gamma_{C}(\mathbf{k}_{A})}\text{ .}%
\end{align}
The $s$-channel expressions are analogous to the $t$-channel. In particular, $P_{s,1}   =\omega_{C}(\mathbf{p}_{C}) + \omega_{A}(\mathbf{p}_{c})-i\frac
{\Gamma_{C}}{2\gamma_{C}(\mathbf{p}_{C})}$ and $P_{s,2} =\omega_{A}(\mathbf{k}_{A})+\omega_{C}(\mathbf{k}_{A}%
)-i\frac{\Gamma_{C}}{2\gamma_{C}(\mathbf{k}_{A})}$.

The differential cross-section in Eq.~\ref{eq:diffcs} can be further simplified
by performing analytically the integral in Eq.~\ref{eq:F1} over the internal
energy $q^{0}$ by appropriately closing `up' or `down' in the complex plane.
In particular, for terms with no $q^{0}$ and for terms with $e^{iq^{0}T+...}$ we
close up (unique pole in $-E_{B}+i\varepsilon$) while for terms proportional
to $e^{-iq^{0}T+...}$ we close down (three poles in $E_{B}+i\varepsilon$ and
in $P_{1,2}$) leading to the expression in Eqs.~\ref{eq:Feq}-\ref{eq:Ifd}.

\bigskip


\begin{thebibliography}{34}%
\makeatletter
\providecommand \@ifxundefined [1]{%
 \@ifx{#1\undefined}
}%
\providecommand \@ifnum [1]{%
 \ifnum #1\expandafter \@firstoftwo
 \else \expandafter \@secondoftwo
 \fi
}%
\providecommand \@ifx [1]{%
 \ifx #1\expandafter \@firstoftwo
 \else \expandafter \@secondoftwo
 \fi
}%
\providecommand \natexlab [1]{#1}%
\providecommand \enquote  [1]{``#1''}%
\providecommand \bibnamefont  [1]{#1}%
\providecommand \bibfnamefont [1]{#1}%
\providecommand \citenamefont [1]{#1}%
\providecommand \href@noop [0]{\@secondoftwo}%
\providecommand \href [0]{\begingroup \@sanitize@url \@href}%
\providecommand \@href[1]{\@@startlink{#1}\@@href}%
\providecommand \@@href[1]{\endgroup#1\@@endlink}%
\providecommand \@sanitize@url [0]{\catcode `\\12\catcode `\$12\catcode
  `\&12\catcode `\#12\catcode `\^12\catcode `\_12\catcode `\%12\relax}%
\providecommand \@@startlink[1]{}%
\providecommand \@@endlink[0]{}%
\providecommand \url  [0]{\begingroup\@sanitize@url \@url }%
\providecommand \@url [1]{\endgroup\@href {#1}{\urlprefix }}%
\providecommand \urlprefix  [0]{URL }%
\providecommand \Eprint [0]{\href }%
\providecommand \doibase [0]{https://doi.org/}%
\providecommand \selectlanguage [0]{\@gobble}%
\providecommand \bibinfo  [0]{\@secondoftwo}%
\providecommand \bibfield  [0]{\@secondoftwo}%
\providecommand \translation [1]{[#1]}%
\providecommand \BibitemOpen [0]{}%
\providecommand \bibitemStop [0]{}%
\providecommand \bibitemNoStop [0]{.\EOS\space}%
\providecommand \EOS [0]{\spacefactor3000\relax}%
\providecommand \BibitemShut  [1]{\csname bibitem#1\endcsname}%
\let\auto@bib@innerbib\@empty
\bibitem [{\citenamefont {Navas}\ \emph {et~al.}(2024)\citenamefont {Navas}
  \emph {et~al.}}]{ParticleDataGroup:2024cfk}%
  \BibitemOpen
  \bibfield  {author} {\bibinfo {author} {\bibfnamefont {S.}~\bibnamefont
  {Navas}} \emph {et~al.} (\bibinfo {collaboration} {Particle Data Group}),\
  }\href {https://doi.org/10.1103/PhysRevD.110.030001} {\bibfield  {journal}
  {\bibinfo  {journal} {Phys. Rev. D}\ }\textbf {\bibinfo {volume} {110}},\
  \bibinfo {pages} {030001} (\bibinfo {year} {2024})}\BibitemShut {NoStop}%
\bibitem [{\citenamefont {Fabbietti}\ \emph {et~al.}(2021)\citenamefont
  {Fabbietti}, \citenamefont {Mantovani~Sarti},\ and\ \citenamefont
  {Vazquez~Doce}}]{Fabbietti:2020bfg}%
  \BibitemOpen
  \bibfield  {author} {\bibinfo {author} {\bibfnamefont {L.}~\bibnamefont
  {Fabbietti}}, \bibinfo {author} {\bibfnamefont {V.}~\bibnamefont
  {Mantovani~Sarti}},\ and\ \bibinfo {author} {\bibfnamefont {O.}~\bibnamefont
  {Vazquez~Doce}},\ }\href {https://doi.org/10.1146/annurev-nucl-102419-034438}
  {\bibfield  {journal} {\bibinfo  {journal} {Ann. Rev. Nucl. Part. Sci.}\
  }\textbf {\bibinfo {volume} {71}},\ \bibinfo {pages} {377} (\bibinfo {year}
  {2021})},\ \Eprint {https://arxiv.org/abs/2012.09806} {arXiv:2012.09806
  [nucl-ex]} \BibitemShut {NoStop}%
\bibitem [{\citenamefont {Peskin}\ and\ \citenamefont
  {Schroeder}(1995)}]{Peskin:1995ev}%
  \BibitemOpen
  \bibfield  {author} {\bibinfo {author} {\bibfnamefont {M.~E.}\ \bibnamefont
  {Peskin}}\ and\ \bibinfo {author} {\bibfnamefont {D.~V.}\ \bibnamefont
  {Schroeder}},\ }\href {https://doi.org/10.1201/9780429503559} {\emph
  {\bibinfo {title} {{An Introduction to quantum field theory}}}}\ (\bibinfo
  {publisher} {Addison-Wesley},\ \bibinfo {address} {Reading, USA},\ \bibinfo
  {year} {1995})\BibitemShut {NoStop}%
\bibitem [{\citenamefont {Peierls}(1961)}]{Peierls:1961zz}%
  \BibitemOpen
  \bibfield  {author} {\bibinfo {author} {\bibfnamefont {R.~F.}\ \bibnamefont
  {Peierls}},\ }\href {https://doi.org/10.1103/PhysRevLett.6.641} {\bibfield
  {journal} {\bibinfo  {journal} {Phys. Rev. Lett.}\ }\textbf {\bibinfo
  {volume} {6}},\ \bibinfo {pages} {641} (\bibinfo {year} {1961})}\BibitemShut
  {NoStop}%
\bibitem [{\citenamefont {Grzadkowski}\ \emph {et~al.}(2022)\citenamefont
  {Grzadkowski}, \citenamefont {Iglicki},\ and\ \citenamefont
  {Mr\'owczy\'nski}}]{Grzadkowski:2021kgi}%
  \BibitemOpen
  \bibfield  {author} {\bibinfo {author} {\bibfnamefont {B.}~\bibnamefont
  {Grzadkowski}}, \bibinfo {author} {\bibfnamefont {M.}~\bibnamefont
  {Iglicki}},\ and\ \bibinfo {author} {\bibfnamefont {S.}~\bibnamefont
  {Mr\'owczy\'nski}},\ }\href {https://doi.org/10.1016/j.nuclphysb.2022.115967}
  {\bibfield  {journal} {\bibinfo  {journal} {Nucl. Phys. B}\ }\textbf
  {\bibinfo {volume} {984}},\ \bibinfo {pages} {115967} (\bibinfo {year}
  {2022})},\ \Eprint {https://arxiv.org/abs/2108.01757} {arXiv:2108.01757
  [hep-ph]} \BibitemShut {NoStop}%
\bibitem [{Note1()}]{Note1}%
  \BibitemOpen
  \bibinfo {note} {Strictly speaking, the exchange channel is of the $u$-type,
  but the literature deals with $t$-channel singularity, see below. The
  scattering angle $\theta $ is chosen as the angle between $\protect \vec
  {p}_C$ and $\protect \vec {k}_A$, in agreement with the
  $t$-channel.}\BibitemShut {Stop}%
\bibitem [{\citenamefont {Melnikov}\ and\ \citenamefont
  {Serbo}(1997)}]{Melnikov:1996iu}%
  \BibitemOpen
  \bibfield  {author} {\bibinfo {author} {\bibfnamefont {K.}~\bibnamefont
  {Melnikov}}\ and\ \bibinfo {author} {\bibfnamefont {V.~G.}\ \bibnamefont
  {Serbo}},\ }\href {https://doi.org/10.1016/S0550-3213(96)00558-5} {\bibfield
  {journal} {\bibinfo  {journal} {Nucl. Phys. B}\ }\textbf {\bibinfo {volume}
  {483}},\ \bibinfo {pages} {67} (\bibinfo {year} {1997})},\ \bibinfo {note}
  {[Erratum: Nucl.Phys.B 662, 409 (2003)]},\ \Eprint
  {https://arxiv.org/abs/hep-ph/9601290} {arXiv:hep-ph/9601290} \BibitemShut
  {NoStop}%
\bibitem [{\citenamefont {Dams}\ and\ \citenamefont
  {Kleiss}(2003)}]{Dams:2002uy}%
  \BibitemOpen
  \bibfield  {author} {\bibinfo {author} {\bibfnamefont {C.}~\bibnamefont
  {Dams}}\ and\ \bibinfo {author} {\bibfnamefont {R.}~\bibnamefont {Kleiss}},\
  }\href {https://doi.org/10.1140/epjc/s2003-01221-6} {\bibfield  {journal}
  {\bibinfo  {journal} {Eur. Phys. J. C}\ }\textbf {\bibinfo {volume} {29}},\
  \bibinfo {pages} {11} (\bibinfo {year} {2003})},\ \Eprint
  {https://arxiv.org/abs/hep-ph/0212301} {arXiv:hep-ph/0212301} \BibitemShut
  {NoStop}%
\bibitem [{\citenamefont {Blinov}\ \emph {et~al.}(1982)\citenamefont {Blinov}
  \emph {et~al.}}]{Blinov:1982vp}%
  \BibitemOpen
  \bibfield  {author} {\bibinfo {author} {\bibfnamefont {A.~E.}\ \bibnamefont
  {Blinov}} \emph {et~al.},\ }\href
  {https://doi.org/10.1016/0370-2693(82)90777-8} {\bibfield  {journal}
  {\bibinfo  {journal} {Phys. Lett. B}\ }\textbf {\bibinfo {volume} {113}},\
  \bibinfo {pages} {423} (\bibinfo {year} {1982})}\BibitemShut {NoStop}%
\bibitem [{\citenamefont {Iglicki}(2023)}]{Iglicki:2022jjf}%
  \BibitemOpen
  \bibfield  {author} {\bibinfo {author} {\bibfnamefont {M.}~\bibnamefont
  {Iglicki}},\ }\href {https://doi.org/10.1007/JHEP06(2023)006} {\bibfield
  {journal} {\bibinfo  {journal} {JHEP}\ }\textbf {\bibinfo {volume} {06}},\
  \bibinfo {pages} {006}},\ \Eprint {https://arxiv.org/abs/2212.00561}
  {arXiv:2212.00561 [hep-ph]} \BibitemShut {NoStop}%
\bibitem [{\citenamefont {Ginzburg}(1996)}]{Ginzburg:1995bc}%
  \BibitemOpen
  \bibfield  {author} {\bibinfo {author} {\bibfnamefont {I.~F.}\ \bibnamefont
  {Ginzburg}},\ }\href {https://doi.org/10.1016/0920-5632(96)00418-5}
  {\bibfield  {journal} {\bibinfo  {journal} {Nucl. Phys. B Proc. Suppl.}\
  }\textbf {\bibinfo {volume} {51}},\ \bibinfo {pages} {85} (\bibinfo {year}
  {1996})},\ \Eprint {https://arxiv.org/abs/hep-ph/9601272}
  {arXiv:hep-ph/9601272} \BibitemShut {NoStop}%
\bibitem [{\citenamefont {Goebel}(1964)}]{Goebel:1964zz}%
  \BibitemOpen
  \bibfield  {author} {\bibinfo {author} {\bibfnamefont {C.}~\bibnamefont
  {Goebel}},\ }\href {https://doi.org/10.1103/PhysRevLett.13.143} {\bibfield
  {journal} {\bibinfo  {journal} {Phys. Rev. Lett.}\ }\textbf {\bibinfo
  {volume} {13}},\ \bibinfo {pages} {143} (\bibinfo {year} {1964})}\BibitemShut
  {NoStop}%
\bibitem [{\citenamefont {Mai}\ \emph {et~al.}(2017)\citenamefont {Mai},
  \citenamefont {Hu}, \citenamefont {Doring}, \citenamefont {Pilloni},\ and\
  \citenamefont {Szczepaniak}}]{Mai:2017vot}%
  \BibitemOpen
  \bibfield  {author} {\bibinfo {author} {\bibfnamefont {M.}~\bibnamefont
  {Mai}}, \bibinfo {author} {\bibfnamefont {B.}~\bibnamefont {Hu}}, \bibinfo
  {author} {\bibfnamefont {M.}~\bibnamefont {Doring}}, \bibinfo {author}
  {\bibfnamefont {A.}~\bibnamefont {Pilloni}},\ and\ \bibinfo {author}
  {\bibfnamefont {A.}~\bibnamefont {Szczepaniak}},\ }\href
  {https://doi.org/10.1140/epja/i2017-12368-4} {\bibfield  {journal} {\bibinfo
  {journal} {Eur. Phys. J. A}\ }\textbf {\bibinfo {volume} {53}},\ \bibinfo
  {pages} {177} (\bibinfo {year} {2017})},\ \Eprint
  {https://arxiv.org/abs/1706.06118} {arXiv:1706.06118 [nucl-th]} \BibitemShut
  {NoStop}%
\bibitem [{\citenamefont {Jackura}\ \emph {et~al.}(2019)\citenamefont
  {Jackura}, \citenamefont {Fern\'andez-Ram\'\i{}rez}, \citenamefont {Mathieu},
  \citenamefont {Mikhasenko}, \citenamefont {Nys}, \citenamefont {Pilloni},
  \citenamefont {Salda\~na}, \citenamefont {Sherrill},\ and\ \citenamefont
  {Szczepaniak}}]{Jackura:2018xnx}%
  \BibitemOpen
  \bibfield  {author} {\bibinfo {author} {\bibfnamefont {A.}~\bibnamefont
  {Jackura}}, \bibinfo {author} {\bibfnamefont {C.}~\bibnamefont
  {Fern\'andez-Ram\'\i{}rez}}, \bibinfo {author} {\bibfnamefont
  {V.}~\bibnamefont {Mathieu}}, \bibinfo {author} {\bibfnamefont
  {M.}~\bibnamefont {Mikhasenko}}, \bibinfo {author} {\bibfnamefont
  {J.}~\bibnamefont {Nys}}, \bibinfo {author} {\bibfnamefont {A.}~\bibnamefont
  {Pilloni}}, \bibinfo {author} {\bibfnamefont {K.}~\bibnamefont {Salda\~na}},
  \bibinfo {author} {\bibfnamefont {N.}~\bibnamefont {Sherrill}},\ and\
  \bibinfo {author} {\bibfnamefont {A.~P.}\ \bibnamefont {Szczepaniak}}
  (\bibinfo {collaboration} {JPAC}),\ }\href
  {https://doi.org/10.1140/epjc/s10052-019-6566-1} {\bibfield  {journal}
  {\bibinfo  {journal} {Eur. Phys. J. C}\ }\textbf {\bibinfo {volume} {79}},\
  \bibinfo {pages} {56} (\bibinfo {year} {2019})},\ \Eprint
  {https://arxiv.org/abs/1809.10523} {arXiv:1809.10523 [hep-ph]} \BibitemShut
  {NoStop}%
\bibitem [{\citenamefont {Mikhasenko}\ \emph {et~al.}(2019)\citenamefont
  {Mikhasenko}, \citenamefont {Wunderlich}, \citenamefont {Jackura},
  \citenamefont {Mathieu}, \citenamefont {Pilloni}, \citenamefont {Ketzer},\
  and\ \citenamefont {Szczepaniak}}]{Mikhasenko:2019vhk}%
  \BibitemOpen
  \bibfield  {author} {\bibinfo {author} {\bibfnamefont {M.}~\bibnamefont
  {Mikhasenko}}, \bibinfo {author} {\bibfnamefont {Y.}~\bibnamefont
  {Wunderlich}}, \bibinfo {author} {\bibfnamefont {A.}~\bibnamefont {Jackura}},
  \bibinfo {author} {\bibfnamefont {V.}~\bibnamefont {Mathieu}}, \bibinfo
  {author} {\bibfnamefont {A.}~\bibnamefont {Pilloni}}, \bibinfo {author}
  {\bibfnamefont {B.}~\bibnamefont {Ketzer}},\ and\ \bibinfo {author}
  {\bibfnamefont {A.~P.}\ \bibnamefont {Szczepaniak}},\ }\href
  {https://doi.org/10.1007/JHEP08(2019)080} {\bibfield  {journal} {\bibinfo
  {journal} {JHEP}\ }\textbf {\bibinfo {volume} {08}},\ \bibinfo {pages}
  {080}},\ \Eprint {https://arxiv.org/abs/1904.11894} {arXiv:1904.11894
  [hep-ph]} \BibitemShut {NoStop}%
\bibitem [{\citenamefont {Nowakowski}\ and\ \citenamefont
  {Pilaftsis}(1993)}]{Nowakowski:1993iu}%
  \BibitemOpen
  \bibfield  {author} {\bibinfo {author} {\bibfnamefont {M.}~\bibnamefont
  {Nowakowski}}\ and\ \bibinfo {author} {\bibfnamefont {A.}~\bibnamefont
  {Pilaftsis}},\ }\href {https://doi.org/10.1007/BF01650437} {\bibfield
  {journal} {\bibinfo  {journal} {Z. Phys. C}\ }\textbf {\bibinfo {volume}
  {60}},\ \bibinfo {pages} {121} (\bibinfo {year} {1993})},\ \Eprint
  {https://arxiv.org/abs/hep-ph/9305321} {arXiv:hep-ph/9305321} \BibitemShut
  {NoStop}%
\bibitem [{\citenamefont {Karamitros}\ and\ \citenamefont
  {Pilaftsis}(2023)}]{Karamitros:2022nnh}%
  \BibitemOpen
  \bibfield  {author} {\bibinfo {author} {\bibfnamefont {D.}~\bibnamefont
  {Karamitros}}\ and\ \bibinfo {author} {\bibfnamefont {A.}~\bibnamefont
  {Pilaftsis}},\ }\href {https://doi.org/10.1103/PhysRevD.108.036007}
  {\bibfield  {journal} {\bibinfo  {journal} {Phys. Rev. D}\ }\textbf {\bibinfo
  {volume} {108}},\ \bibinfo {pages} {036007} (\bibinfo {year} {2023})},\
  \Eprint {https://arxiv.org/abs/2208.10425} {arXiv:2208.10425 [hep-th]}
  \BibitemShut {NoStop}%
\bibitem [{Note2()}]{Note2}%
  \BibitemOpen
  \bibinfo {note} {In Ref.~\cite {Grzadkowski:2021kgi} the width of the state
  $B$ is generated by medium effects. The latter being {\protect \it de facto}
  always present, the $t$-channel singularity is absent if the scattering
  occurs in the medium. In this respect, this type of solution is complete, but
  requires an external medium/field. Removing it, even in principle, restores
  the singularity.}\BibitemShut {Stop}%
\bibitem [{Note3()}]{Note3}%
  \BibitemOpen
  \bibinfo {note} {Namely, one may cut $AAB\rightarrow AAB$ diagram through an
  internal $B$ ({\protect \it c.f.} Fig.~\ref {fig:3to3}), resulting in
  distinct sub-processes of the $AB\rightarrow AB$ types. A full treatment of
  this case is left for the future.}\BibitemShut {Stop}%
\bibitem [{\citenamefont {Collins}(2019)}]{Collins:2019ozc}%
  \BibitemOpen
  \bibfield  {author} {\bibinfo {author} {\bibfnamefont {J.}~\bibnamefont
  {Collins}},\ }\Eprint {https://arxiv.org/abs/1904.10923} {arXiv:1904.10923
  [hep-ph]}  (\bibinfo {year} {2019})\BibitemShut {NoStop}%
\bibitem [{\citenamefont {Bernardini}\ \emph {et~al.}(1993)\citenamefont
  {Bernardini}, \citenamefont {Maiani},\ and\ \citenamefont
  {Testa}}]{Bernardini:1993qj}%
  \BibitemOpen
  \bibfield  {author} {\bibinfo {author} {\bibfnamefont {C.}~\bibnamefont
  {Bernardini}}, \bibinfo {author} {\bibfnamefont {L.}~\bibnamefont {Maiani}},\
  and\ \bibinfo {author} {\bibfnamefont {M.}~\bibnamefont {Testa}},\ }\href
  {https://doi.org/10.1103/PhysRevLett.71.2687} {\bibfield  {journal} {\bibinfo
   {journal} {Phys. Rev. Lett.}\ }\textbf {\bibinfo {volume} {71}},\ \bibinfo
  {pages} {2687} (\bibinfo {year} {1993})}\BibitemShut {NoStop}%
\bibitem [{\citenamefont {Blasone}\ \emph {et~al.}(2023)\citenamefont
  {Blasone}, \citenamefont {Giacosa}, \citenamefont {Smaldone},\ and\
  \citenamefont {Torrieri}}]{Blasone:2023brf}%
  \BibitemOpen
  \bibfield  {author} {\bibinfo {author} {\bibfnamefont {M.}~\bibnamefont
  {Blasone}}, \bibinfo {author} {\bibfnamefont {F.}~\bibnamefont {Giacosa}},
  \bibinfo {author} {\bibfnamefont {L.}~\bibnamefont {Smaldone}},\ and\
  \bibinfo {author} {\bibfnamefont {G.}~\bibnamefont {Torrieri}},\ }\href
  {https://doi.org/10.1140/epjc/s10052-023-11867-3} {\bibfield  {journal}
  {\bibinfo  {journal} {Eur. Phys. J. C}\ }\textbf {\bibinfo {volume} {83}},\
  \bibinfo {pages} {736} (\bibinfo {year} {2023})},\ \Eprint
  {https://arxiv.org/abs/2305.07107} {arXiv:2305.07107 [hep-ph]} \BibitemShut
  {NoStop}%
\bibitem [{\citenamefont {Blasone}\ \emph {et~al.}(2025)\citenamefont
  {Blasone}, \citenamefont {Giacosa}, \citenamefont {Smaldone},\ and\
  \citenamefont {Torrieri}}]{Blasone:2025hjw}%
  \BibitemOpen
  \bibfield  {author} {\bibinfo {author} {\bibfnamefont {M.}~\bibnamefont
  {Blasone}}, \bibinfo {author} {\bibfnamefont {F.}~\bibnamefont {Giacosa}},
  \bibinfo {author} {\bibfnamefont {L.}~\bibnamefont {Smaldone}},\ and\
  \bibinfo {author} {\bibfnamefont {G.}~\bibnamefont {Torrieri}},\ }\href
  {https://doi.org/10.1140/epjc/s10052-025-14165-2} {\bibfield  {journal}
  {\bibinfo  {journal} {Eur. Phys. J. C}\ }\textbf {\bibinfo {volume} {85}},\
  \bibinfo {pages} {523} (\bibinfo {year} {2025})},\ \Eprint
  {https://arxiv.org/abs/2501.17111} {arXiv:2501.17111 [hep-ph]} \BibitemShut
  {NoStop}%
\bibitem [{\citenamefont {Giacosa}(2016)}]{Giacosa:2015mpm}%
  \BibitemOpen
  \bibfield  {author} {\bibinfo {author} {\bibfnamefont {F.}~\bibnamefont
  {Giacosa}},\ }\href {https://doi.org/10.5506/APhysPolB.47.2135} {\bibfield
  {journal} {\bibinfo  {journal} {Acta Phys. Polon. B}\ }\textbf {\bibinfo
  {volume} {47}},\ \bibinfo {pages} {2135} (\bibinfo {year} {2016})},\ \Eprint
  {https://arxiv.org/abs/1512.00232} {arXiv:1512.00232 [hep-ph]} \BibitemShut
  {NoStop}%
\bibitem [{\citenamefont {Gavassino}\ and\ \citenamefont
  {Giacosa}(2022)}]{Gavassino:2022ksl}%
  \BibitemOpen
  \bibfield  {author} {\bibinfo {author} {\bibfnamefont {L.}~\bibnamefont
  {Gavassino}}\ and\ \bibinfo {author} {\bibfnamefont {F.}~\bibnamefont
  {Giacosa}},\ }\href {https://doi.org/10.1103/PhysRevA.106.042215} {\bibfield
  {journal} {\bibinfo  {journal} {Phys. Rev. A}\ }\textbf {\bibinfo {volume}
  {106}},\ \bibinfo {pages} {042215} (\bibinfo {year} {2022})},\ \Eprint
  {https://arxiv.org/abs/2206.05125} {arXiv:2206.05125 [physics.gen-ph]}
  \BibitemShut {NoStop}%
\bibitem [{\citenamefont {Civitarese}\ and\ \citenamefont
  {Gadella}(2004)}]{Civitarese:2004xbt}%
  \BibitemOpen
  \bibfield  {author} {\bibinfo {author} {\bibfnamefont {O.}~\bibnamefont
  {Civitarese}}\ and\ \bibinfo {author} {\bibfnamefont {M.}~\bibnamefont
  {Gadella}},\ }\href {https://doi.org/10.1016/j.physrep.2004.03.001}
  {\bibfield  {journal} {\bibinfo  {journal} {Phys. Rept.}\ }\textbf {\bibinfo
  {volume} {396}},\ \bibinfo {pages} {41} (\bibinfo {year} {2004})}\BibitemShut
  {NoStop}%
\bibitem [{\citenamefont {de~la Madrid}\ and\ \citenamefont
  {Gadella}(2002)}]{delaMadrid:2002cz}%
  \BibitemOpen
  \bibfield  {author} {\bibinfo {author} {\bibfnamefont {R.}~\bibnamefont
  {de~la Madrid}}\ and\ \bibinfo {author} {\bibfnamefont {M.}~\bibnamefont
  {Gadella}},\ }\href {https://doi.org/10.1119/1.1466817} {\bibfield  {journal}
  {\bibinfo  {journal} {Am. J. Phys.}\ }\textbf {\bibinfo {volume} {70}},\
  \bibinfo {pages} {626} (\bibinfo {year} {2002})},\ \Eprint
  {https://arxiv.org/abs/quant-ph/0201091} {arXiv:quant-ph/0201091}
  \BibitemShut {NoStop}%
\bibitem [{\citenamefont {Gegelia}\ and\ \citenamefont
  {Scherer}(2010)}]{Gegelia:2010nmt}%
  \BibitemOpen
  \bibfield  {author} {\bibinfo {author} {\bibfnamefont {J.}~\bibnamefont
  {Gegelia}}\ and\ \bibinfo {author} {\bibfnamefont {S.}~\bibnamefont
  {Scherer}},\ }\href {https://doi.org/10.1140/epja/i2010-10955-5} {\bibfield
  {journal} {\bibinfo  {journal} {Eur. Phys. J. A}\ }\textbf {\bibinfo {volume}
  {44}},\ \bibinfo {pages} {425} (\bibinfo {year} {2010})},\ \Eprint
  {https://arxiv.org/abs/0910.4280} {arXiv:0910.4280 [hep-ph]} \BibitemShut
  {NoStop}%
\bibitem [{\citenamefont {Potapov}\ and\ \citenamefont
  {Taylor}(1977{\natexlab{a}})}]{Potapov:1977ux}%
  \BibitemOpen
  \bibfield  {author} {\bibinfo {author} {\bibfnamefont {V.~S.}\ \bibnamefont
  {Potapov}}\ and\ \bibinfo {author} {\bibfnamefont {J.~R.}\ \bibnamefont
  {Taylor}},\ }\href@noop {} {\bibfield  {journal} {\bibinfo  {journal} {Phys.
  Rev. A}\ }\textbf {\bibinfo {volume} {16}} (\bibinfo {year}
  {1977}{\natexlab{a}})}\BibitemShut {NoStop}%
\bibitem [{\citenamefont {Potapov}\ and\ \citenamefont
  {Taylor}(1977{\natexlab{b}})}]{Potapov:1977sr}%
  \BibitemOpen
  \bibfield  {author} {\bibinfo {author} {\bibfnamefont {V.~S.}\ \bibnamefont
  {Potapov}}\ and\ \bibinfo {author} {\bibfnamefont {J.~R.}\ \bibnamefont
  {Taylor}},\ }\href@noop {} {\bibfield  {journal} {\bibinfo  {journal} {Phys.
  Rev. A}\ }\textbf {\bibinfo {volume} {16}} (\bibinfo {year}
  {1977}{\natexlab{b}})}\BibitemShut {NoStop}%
\bibitem [{\citenamefont {Rubin}\ \emph {et~al.}(1966)\citenamefont {Rubin},
  \citenamefont {Sugar},\ and\ \citenamefont {Tiktopoulos}}]{Rubin:1966zz}%
  \BibitemOpen
  \bibfield  {author} {\bibinfo {author} {\bibfnamefont {M.}~\bibnamefont
  {Rubin}}, \bibinfo {author} {\bibfnamefont {R.}~\bibnamefont {Sugar}},\ and\
  \bibinfo {author} {\bibfnamefont {G.}~\bibnamefont {Tiktopoulos}},\ }\href
  {https://doi.org/10.1103/PhysRev.146.1130} {\bibfield  {journal} {\bibinfo
  {journal} {Phys. Rev.}\ }\textbf {\bibinfo {volume} {146}},\ \bibinfo {pages}
  {1130} (\bibinfo {year} {1966})}\BibitemShut {NoStop}%
\bibitem [{Note4()}]{Note4}%
  \BibitemOpen
  \bibinfo {note} {A universal definition of an analogous of the cross-section
  for three-body scattering is not feasible, since the specific invariant flux
  does not exist. Physically, for three particles to interact, all three
  worldlines must meet in a very small space--time region. The frequency of
  such coincidences depends on particle densities, correlations, and beam
  geometry}\BibitemShut {NoStop}%
\bibitem [{\citenamefont {Hansen}\ and\ \citenamefont
  {Sharpe}(2014)}]{Hansen:2014eka}%
  \BibitemOpen
  \bibfield  {author} {\bibinfo {author} {\bibfnamefont {M.~T.}\ \bibnamefont
  {Hansen}}\ and\ \bibinfo {author} {\bibfnamefont {S.~R.}\ \bibnamefont
  {Sharpe}},\ }\href {https://doi.org/10.1103/PhysRevD.90.116003} {\bibfield
  {journal} {\bibinfo  {journal} {Phys. Rev. D}\ }\textbf {\bibinfo {volume}
  {90}},\ \bibinfo {pages} {116003} (\bibinfo {year} {2014})},\ \Eprint
  {https://arxiv.org/abs/1408.5933} {arXiv:1408.5933 [hep-lat]} \BibitemShut
  {NoStop}%
\bibitem [{\citenamefont {Baeza-Ballesteros}\ \emph {et~al.}(2023)\citenamefont
  {Baeza-Ballesteros}, \citenamefont {Bijnens}, \citenamefont {Husek},
  \citenamefont {Romero-L{\'o}pez}, \citenamefont {Sharpe},\ and\ \citenamefont
  {Sj{\"o}}}]{Baeza-Ballesteros:2023ljl}%
  \BibitemOpen
  \bibfield  {author} {\bibinfo {author} {\bibfnamefont {J.}~\bibnamefont
  {Baeza-Ballesteros}}, \bibinfo {author} {\bibfnamefont {J.}~\bibnamefont
  {Bijnens}}, \bibinfo {author} {\bibfnamefont {T.}~\bibnamefont {Husek}},
  \bibinfo {author} {\bibfnamefont {F.}~\bibnamefont {Romero-L{\'o}pez}},
  \bibinfo {author} {\bibfnamefont {S.~R.}\ \bibnamefont {Sharpe}},\ and\
  \bibinfo {author} {\bibfnamefont {M.}~\bibnamefont {Sj{\"o}}},\ }\href
  {https://doi.org/10.1007/JHEP05(2023)187} {\bibfield  {journal} {\bibinfo
  {journal} {JHEP}\ }\textbf {\bibinfo {volume} {05}},\ \bibinfo {pages}
  {187}},\ \Eprint {https://arxiv.org/abs/2303.13206} {arXiv:2303.13206
  [hep-ph]} \BibitemShut {NoStop}%
\end{thebibliography}

%

\end{document}